\documentclass[11pt,twoside]{article}

\usepackage{asp2006}
\usepackage{epsf}
\usepackage{psfig}
\usepackage{lscape}
\usepackage{graphicx}

\begin{document}
\title{Frequency shifts of the individual low-degree p modes during solar cycle~23 and its extended minimum}   
\author{D. Salabert$^{1,2}$, R.~A. Garc\'ia$^3$,  P.~L. Pall\'e$^{1,2}$, and S.~J. Jim\'enez-Reyes$^{1}$}   

\affil{$^{1}$Instituto de Astrof\'isica de Canarias,  E-38200 La Laguna, Tenerife, Spain}   
\affil{$^{2}$Departamento de Astrof\'isica, Universidad de La Laguna, E-38205 La Laguna, Tenerife, Spain}
\affil{$^{3}$Laboratoire AIM, CEA/DSM-CNRS, Universit\'e Paris 7 Diderot, IRFU/SAp, Centre de Saclay, F-91191 Gif-sur-Yvette, France}

\begin{abstract} 
We study the response of the low-degree solar p-mode frequencies to the unusual extension of the minimum of solar surface activity since 2007. Helioseismic observations collected by the space-based, Sun-as-a-star GOLF instrument and by the ground-based, multi-site network GONG (integrated signal) are analyzed. Temporal variations of the low-degree ($l=0,1,2$), p-mode frequencies are obtained. Although the known correlation of the frequency changes with the solar surface activity is recovered for the period 1996--2007, since the second half of  2007 and until July 2009 (latest period analyzed) we notice a peculiar behavior amongst modes of different angular degrees. In particular, a clear increase of the $l=0$ and $l=2$ p-mode frequencies is consistently obtained since late 2007, while  the $l=1$ frequencies follow the general decreasing trend of surface activity. We interpret these differences in the frequency shifts of individual low-degree modes as indicative of variations at high latitudes in the magnetic flux beneath the surface of the Sun related to the onset of solar cycle~24.

\end{abstract}

\section{Introduction}
Temporal variations of the low-degree (low-$l$), solar p-mode frequencies with solar activity were first reported by \citet{wood85}, who found that the $l = 0$ and $l=1$ mode frequencies in the 5-min band decreased by $\sim 0.42$~$\mu$Hz between 1980 (near solar maximum) and 1984 (near solar minimum). These early observations were later on confirmed by \citet{palle89} using data spanning the entire solar cycle 21 (1977-1988). As higher quality and continuous helioseismic data became available, more detailed analysis were carried out revealing the highly-correlated sensitivity of the solar oscillation acoustic frequencies to the solar surface activity at low-  \citep{chaplin01,gelly02,salabert04} and high-angular degrees \citep{chano01,howe02,salabert06a}. Angular-degree dependence of the frequency shifts  at intermediate and high $l$ was also observed \citep{chano01}. Moreover, \citet{howe02} showed close temporal and spatial correlation of  the latitude distribution of the high-degree shifts with the surface magnetic field. Marginal but still significant degree dependence of the low-degree frequency shifts was uncovered as well \citep{chano04b,chaplin04}. 
However, the origin of the frequency shifts is far from being properly understood. The form and the degree dependence of the shifts would favor near-surface phenomenon but however they cannot be purely explained by structural changes: the magnetic field is somehow involved in the mechanism.
As the solar oscillation frequencies, the p-mode amplitudes and linewidths, for instance, were also proven to be sensitive to the solar activity cycle in both Sun-as-a-star \citep{chaplin00,salabert03,chano04a} and spatially-resolved \citep{komm00,salabert06b} observations. 

The frequency shifts being closely correlated with solar surface activity proxies during the past solar cycles \citep{broom09}, the response of the solar oscillations to the current unusually long and deep solar activity minimum is of particular interest. By analyzing 4768 days collected by the space-based helioseismic instrument Global Oscillations at Low Frequency (GOLF) instrument  \citep{gabriel95} onboard the {\it Solar and Heliospheric Observatory (SOHO)} spacecraft, \citet{salabert09} observed that the frequency shifts of the  $l=0$ and $l=2$ modes show a sharp rise from the end of 2007, while no significant surface activity is visible on the Sun. On the other hand, the $l = 1$ modes follow the general decreasing trend of solar surface activity. The differences between individual angular degrees can be interpreted as different geometrical responses to the spatial distribution of the solar magnetic field beneath the surface of the Sun, indicating variations in the magnetic flux at high latitudes related to the onset of solar cycle~24.
Significant variations of the p-mode frequencies during the current minimum in contrast to the surface activity observations over the same period were also reported by \citet{broom09}. However, they did not analyzed the variations in term of individual angular degree. Furthermore, \citet{howe09} showed that the lack of sunspots and the low-activity levels during the current minimum can be explained by a slower than usual jet stream associated with the production of sunspots. These streams originating from the poles every 11 years migrate slowly below the surface towards the equator. 

We present here an updated analysis of the GOLF observations until July 2009, allowing us to include an extra measurement with a 100\% filling factor in comparison with \citet{salabert09}. We also present results obtained by performing the same analysis on the integrated time series of the Global Oscillation Network Group \citep[GONG;][]{harvey96}. We discuss the temporal dependence of the frequency shifts at individual angular degree and their behaviors during the current extended minimum of activity.

\section{Observations and analysis}
\subsection{Data sets}
A total of 4836 days of velocity GOLF time series \citep{ulrich00,garcia05} calibrated following the method detailed in \citet{chano03} were analyzed. This dataset, starting on 1996 April 11 and ending on 2009 July 8, was split into contiguous 365-day subseries. Each series was allowed to overlap by 91.25 days (i.e., a four-time overlap), resulting in a total of 50 non-independent time series. The mean duty cycle during this period was 95.0\%.

We also analyzed 5110 days of the integrated time series of the ground-based, multi-site GONG project, spanning the period from 1995 May 7 to 2009 May 3. The mean duty cycle was 85.4\%. As for the GOLF observations, the GONG data were split into 365-day subseries with a four-time overlap, resulting in a total of 52  non-independent time series.
 
\subsection{Calculation of the frequency temporal variations}
 The power spectrum of each 365-day time series was fitted in order to estimate the mode parameters. The fitting was performed by using a multi-step iterative method \citep{gelly02,salabert07}, in which each mode component is parameterized by a modified Lorentzian model. The mode parameters were extracted using a standard likelihood maximization function following the $\chi^2$ with 2 degrees-of-freedom statistics of the power spectrum. The formal uncertainties in each mode parameter were derived from the inverse Hessian matrix. Details of the fitting procedure can be found in \citet{salabert09}.
Due to the spatial resolution of the original GONG data, part of the power from the  higher angular degrees ($l \geq 4$) are present in the integrated GONG signal. These leaks were taken into account during the fitting by including information from the GONG leakage matrix \citep{hill98}.

The temporal variations of the p-mode frequencies were calculated by comparing each fitted frequency with a reference, taken by averaging the frequencies over the entire set of analyzed spectra for each data sets. The formal errors returned by the fits were used as weights. The frequency shifts were then defined as the difference between these reference values and the frequencies of the corresponding modes observed at different dates. The weighted averages of these frequency shifts were then calculated between 2000 and 3300~$\mu$Hz.
Mean values of daily measurements of the 10.7-cm radio flux were obtained over the same 365-day subseries and used as a proxy of the solar surface activity.

\section{Results}
\begin{figure}
\centering
\includegraphics[width=0.51\textwidth,angle=90]{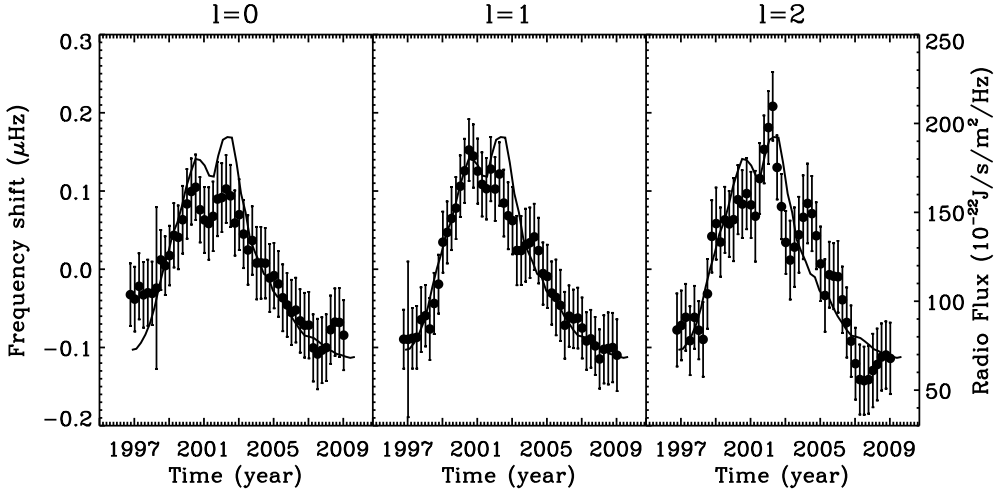}
\caption{Frequency shifts of the $l=0,1$, and 2 solar p modes (left to right panels) extracted from the analysis of the 365-day GOLF spectra. The associated error bars are also represented. The corresponding 10.7-cm radio flux averaged over the same 365-day timespan is shown as a proxy of the solar surface activity (solid line).}
\label{fig:fshift_golf}
\end{figure}

\begin{figure}
\centering
\includegraphics[width=0.51\textwidth,angle=90]{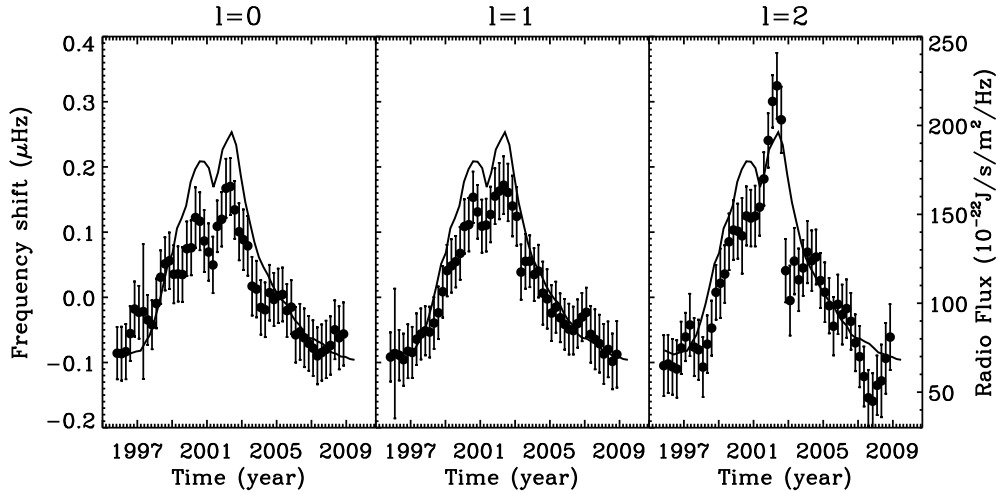}
\caption{Frequency shifts of the $l=0,1$, and 2 solar p modes (left to right panels) extracted from the analysis of the 365-day integrated GONG spectra. The associated error bars are also represented. The corresponding 10.7-cm radio flux averaged over the same 365-day timespan is shown as a proxy of the solar surface activity (solid line). Note the different scale on the $y$ axis compared to Fig.~\ref{fig:fshift_golf}.}
\label{fig:fshift_gong}
\end{figure}

Figures~\ref{fig:fshift_golf} and \ref{fig:fshift_gong} show the temporal dependence of the  frequency shifts at each angular degree $l=0,1$, and 2, measured from the GOLF and from the integrated GONG 365-day spectra respectively. As observed by \citet{salabert09}, the frequency shifts at $l = 0$ and $ l =2$ show a sharp rise from the end of 2007, which is quite pronounced at $l = 2$. At $l = 1$, the frequency shifts keep decreasing following the general trend of solar surface activity. The results are consistent between the space-based GOLF and the ground-based, integrated GONG instruments.

The differences in the frequency shifts of individual low-$l$ modes may be interpreted as different geometrical responses to the spatial distribution of the solar magnetic field beneath the solar surface. Indeed, oscillation modes with lower $l$ values have different sensitivities to different latitudes \citep{chano04b}. In the GOLF and integrated GONG data, the observed $l=1$ frequency is a weighted measurement of the visible components ($l=1, |m|=1$), while the $l=2$ frequency corresponds to the weighted measurement of the zonal ($l=2, m=0$) and sectoral ($l=2, |m|=2$) components\footnote{Because the rotation axis of the Sun lies close to the plane of the sky from the observing point,  full-disk observations, such as GOLF and integrated GONG data, are only sensitive to the modes with $l+|m|$ even.}. The sectoral modes are more concentrated along the equator, while the zonal modes are most sensitive to the high latitudes. Thus, the $l=2$ modes as seen by full-disk instruments are more sensitive to the high latitudes of the Sun. They have larger frequency variations than the $l=0$ modes, which are averaged across the entire visible solar disk, while the full-disk $l=1$ mode is more concentrated along the equator.

\section{Conclusions}
We analyzed integrated helioseismic observations collected by the space-based GOLF and ground-based, multi-site network GONG (integrated signal) instruments, and studied the response of the low-degree p-mode frequencies to the unusually extended and deep surface activity minimum of solar cycle~23.
While closely correlated with the surface activity proxies during the past solar cycles, the temporal variations of the individual $l=0$ and $l=2$ mode frequencies (full-disk), which are more sensitive to higher latitudes, show an upturn from the end of 2007, when no significant activity is observable on the surface of the Sun.  On the other hand, the variations of the $l=1$ mode frequencies (full-disk), which are less sensitive to higher latitudes, show no evidence of an upturn and follow the decreasing trend of surface activity. Similar results were obtained with the GOLF and the integrated GONG data sets. These differences between individual angular degrees may indicate that the magnetic effects related to the new solar cycle~24 happening beneath the surface  and responsible for the frequency shifts have started in late 2007 at high latitudes.

\acknowledgements 
The GOLF instrument onboard SOHO is a cooperative effort of many individuals, to whom we are indebted. SOHO is a project of international collaboration between ESA and NASA. This work utilizes data obtained by the Global Oscillation Network Group (GONG) program, managed by the National Solar Observatory, which is operated by AURA, Inc. under a cooperative agreement with the National Science Foundation. The data were acquired by instruments operated by the Big Bear Solar Observatory, High Altitude Observatory,
Learmonth Solar Observatory, Udaipur Solar Observatory, Instituto de Astrof\'{\i}sica de Canarias, and Cerro Tololo Interamerican Observatory. The 10.7-cm radio flux data were obtained from the National Geophysical Data Center. D.S. acknowledges the support of the grant PNAyA2007-62650 from the Spanish National Research Plan. This work was supported by the European Helio- and Asteroseismology Network (HELAS), a major international collaboration funded by the European Commission's FP6, and by the CNES/GOLF grant at the SAp/CEA-Saclay.

\end{document}